\documentclass[a4paper,final]{appolb}
\usepackage{graphicx}
\usepackage{amsmath,amssymb}

\newcommand{\ket}[1]{| {#1} \rangle}
\newcommand{\bra}[1]{\langle {#1} |}
\newcommand{\inproduct}[2]{\langle {#1} | {#2} \rangle}

\begin{document}
\title{Self-consistent description of nuclear photoabsorption cross sections
\thanks{Presented at Zakopane Conference on Nuclear Physics, Zakopane,
Poland, August 30--September 5, 2010}%
}
\author{Takashi~Nakatsukasa$^{1,2}$,
Paolo~Avogadro$^{1}$,
Shuichiro~Ebata$^{1,3}$,
Tsunenori~Inakura$^{1}$,
Kenichi~Yoshida$^{1}$
\address{$^{1}$RIKEN Nishina Center, Wako 351-0198, Japan}
\address{$^{2}$Center for Computational Sciences, University of Tsukuba,
Tsukuba 305-8571, Japan}
\address{$^{3}$Graduate School of Pure and Applied Sciences,
University of Tsukuba, Tsukuba 305-8571, Japan}
}
\maketitle
\begin{abstract}
Several approaches to photonuclear
reactions, based on the time-dependent density-functional theory,
have been developed recently.
The standard linearization leads to the random-phase
approximation (RPA) or the quasiparticle-random-phase approximation (QRPA).
We have developed a parallelized QRPA computer program for axially
deformed nuclei.
We also present a feasible approach to the (Q)RPA calculation,
that is the finite amplitude method (FAM).
We show results of photoabsorption cross sections for deformed nuclei
using the QRPA and FAM calculations.
Finally, the canonical-basis approach to the
time-dependent Hartree-Fock-Bogoliubov method is presented,
to demonstrate its feasibility and usefulness.
\end{abstract}
\PACS{21.10.Re; 21.60.Jz; 24.30.Cz}
  
\section{Introduction}
Photonuclear reaction cross sections are the fundamental
properties in nuclear systems.
In the energy region of giant resonances ($E=10\sim 30$ MeV), 
the absorption process is dominated by the electric dipole excitations.
The giant dipole resonance (GDR) has been of significant interest in
studies of nuclear structure and reaction.
It exhausts almost 100 \% of the energy-weighted sum-rule value,
corresponding to a collective oscillation of neutrons against the
protons.
A typical measurement of the GDR in stable nuclei is the photoneutron
cross section measurement using monoenergetic photons \cite{BF75}.
The energy of the GDR peak was found to have a mass dependence
midway between $A^{-1/3}$ and $A^{-1/6}$ which
correspond to the Steinwedel-Jensen and Goldhaber-Teller models, respectively
\cite{RS80}.
The energy-weighted sum-rule value is larger than the classical
Thomas-Reiche-Kuhn (TRK) value by 20 \% \cite{BF75} in average.
The general trend of the width of the GDR is well correlated with
the neutron magic numbers, which may suggest that the main origin of the
spreading width is the shape fluctuations in the ground state \cite{Car74,BM75}.
The double-peak structure in GDR appears for
axially deformed nuclei,  known as the deformation splitting,
because of the different frequencies for vibrations
along and perpendicular to the symmetry axis \cite{BF75}.

In a microscopic point of view, one can construct the giant resonance
from a superposition of particle-hole excitations.
Since dynamics of the giant resonances are basically in a small-amplitude
regime, the random-phase approximation (RPA) \cite{RS80}
has been extensively utilized for studies of their properties.
Although the spreading width $\Gamma^\downarrow$ is not taken into
account in the RPA level, main features of the giant resonance are well
reproduced.
In this paper, we present three theoretical approaches to studies of
the nuclear response;
the standard quasiparticle RPA (QRPA) \cite{RS80},
the finite amplitude method (FAM) \cite{NIY07},
and the canonical-basis time-dependent Hartree-Fock-Bogoliubov
(Cb-TDHFB) method \cite{Eba10}.
The numerical results will be shown, mainly focused on
the photoabsorption cross section.

\section{Quasiparticle random-phase approximation for axially deformed nuclei}
\label{sec:QRPA}
The quasiparticle RPA (QRPA) is a standard method to calculate linear
response in heavy open-shell nuclei \cite{RS80}.
However, since its application to realistic energy functionals requires
a large computational task and a complicated programing,
the QRPA calculation for heavy deformed nuclei is still a challenging
subject at present.

We have recently developed a parallelized computer code of the QRPA
based on the Hartree-Fock-Bogoliubov (HFB) state with the
Skyrme functionals,
which is an extended version of that developed in Ref. \cite{YG08-2},
to include the residual spin-orbit interaction.
A missing part is only the residual Coulomb interaction that does not
significantly affect nuclear response functions
(See Sec.~\ref{sec:Cb-TDHFB-cal}).

First, we solve the following self-consistent HFB equation for
the quasiparticle states:
\begin{equation}
\label{HFB}
\begin{pmatrix}
h-\lambda & \Delta \\
-\Delta^*& -(h-\lambda)^*
\end{pmatrix}
\begin{pmatrix}
U_\mu \\
V_\mu 
\end{pmatrix}
=E_\mu
\begin{pmatrix}
U_\mu \\
V_\mu 
\end{pmatrix}
\end{equation}
where the single-particle Hamiltonian $h[\rho,\kappa]$ and the pair potential
$\Delta[\rho,\kappa]$ are functionals of the density $\rho$ and
the pairing tensor $\kappa$.
The self-consistent solution of
Eq. (\ref{HFB}) determines the ground-state densities $(\rho_0,\kappa_0)$
and the ground-state Hamiltonians
$(h_0,\Delta_0)$.
To describe the nuclear deformation
and the pairing correlations, simultaneously, in good account of the continuum,
we solve the HFB equations
in the cylindrical coordinate space.
We assume axial and reflection symmetries in the ground state.
To reduce the QRPA-matrix dimension,
we introduce a cut-off energy $E^{\rm 2qp}_{\rm c}=60$ MeV
for the two-quasiparticle states.
For instance,
the number of two-quasiparticle states becomes
about 38,000 for the $K^{\pi}=0^{-}$ excitation in $^{154}$Sm.
Then, we calculate the QRPA matrix elements and diagonalize the matrix,
to obtain the QRPA normal modes.
\begin{equation}
\sum_{\gamma\delta}
\begin{pmatrix}
A_{\alpha\beta,\gamma\delta} & B_{\alpha\beta,\gamma\delta} \\
-B_{\alpha\beta,\gamma\delta} & -A_{\alpha\beta,\gamma\delta}
\end{pmatrix}
\begin{pmatrix}
X_{\gamma\delta} \\
Y_{\gamma\delta}
\end{pmatrix}
= \hbar\omega
\begin{pmatrix}
X_{\gamma\delta} \\
Y_{\gamma\delta}
\end{pmatrix}
\end{equation}
Since the spreading effect is missing in this calculation,
the dipole strength of each discrete eigenmode is
folded by the Lorentzian curve with a smoothing parameter $\Gamma$.
A more detailed description can be found in Ref. \cite{YN10}.

\begin{figure}[t]
\begin{center}
\includegraphics[width=0.7\textwidth]{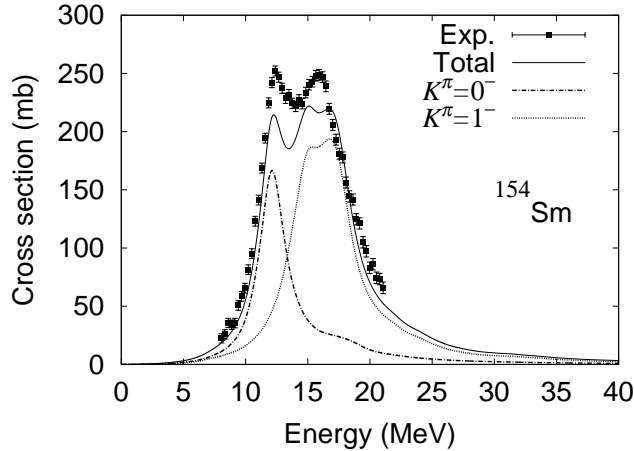}
\caption{Calculated (lines) and experimental (symbols) photoabsorption cross
section in $^{154}$Sm.
The SkM* parameter set and
the smoothing parameter of $\Gamma=2$ MeV is used.
See text for details.
Experimental values are taken from Ref. \cite{Car74}.
}
\label{fig:154Sm_QRPA}
\end{center}
\end{figure}
We show in Fig.~\ref{fig:154Sm_QRPA} the photoabsorption
cross section for $^{154}$Sm.
The HFB calculation with the SkM* parameters produces the ground state
in a prolate deformation of $\beta=0.31$.
It clearly shows a deformation splitting due to a prolate deformation
of the ground state.
The experimental data \cite{Car74} are well reproduced in the calculation.
We have carried out a systematic analysis on Nd and Sm isotopes and
have found that the spreading effect with $\Gamma=2$ MeV can
well reproduce experimental data from spherical, transitional,
to deformed nuclei \cite{YN10}.
Especially, the agreement on the evolution of the GDR width as a function
of the neutron number is excellent.

\section{Finite amplitude method}
\label{sec:FAM}
In this section, we recapitulate the methodology of
the finite amplitude method (FAM) we have developed for
small-amplitude oscillations based on the time-dependent density-functional
theory \cite{NIY07,INY09}.

\subsection{FAM without pairing correlations}
\label{sec:FAM_HF}
First, we discuss the case that the energy density functional is
represented by normal density $\rho$ only.
In this case, the density can be expressed by the Kohn-Sham orbitals,
$\rho =  \sum_i | \phi_i \rangle \langle \phi_i |$,
where the subscript $i$ indicates the occupied orbitals
($i=1,2,\cdots,A$).
The linear-response equation to a weak external field with a fixed
frequency, $V_\mathrm{ext}(\omega)$, can be expressed in terms of
the forward and backward amplitudes, $\ket{X_i(\omega)}$ and
$\bra{Y_i(\omega)}$.
\begin{eqnarray}
\omega \, | X_i(\omega) \rangle = \left( h_0 - \epsilon_i \right) | X_i(\omega) \rangle
          + \hat{P} \left\{ V_\mathrm{ext}(\omega) + \delta h(\omega) \right\} | \phi_i \rangle  , \label{RPAeqX}\\
- \omega \, \langle Y_i(\omega) | = \langle Y_i(\omega) | \left( h_0 - \epsilon_i \right)
          + \langle \phi_i | \left\{ V_\mathrm{ext}(\omega) + \delta h(\omega) \right\} \hat{P}  . \label{RPAeqY}
\end{eqnarray}
where the operator $\hat{P}$ denotes the projector
onto the particles space, $\hat{P} =  1 -  \sum_i | \phi_i \rangle
\langle \phi_i |$.
Usually, the residual field $\delta h(\omega)$ is
expanded to the first order with respect to
$| X_i(\omega)\rangle$ and
$| Y_i(\omega)\rangle$.
This leads to the well-known matrix form of
the linear-response equation, known as the RPA.
For deformed nuclei,
the calculation of these matrix
elements is time-consuming in practice and
their storage requires a large memory capacity.
In the FAM, we do not explicitly linearize the equations.
Instead, we utilize the
fact that the linearization can be numerically achieved for
$\delta h(\omega) = h[\rho_0 + \delta\rho(\omega)] - h_0$,
if the transition density $\delta\rho(\omega)$ is small enough to
validate the linear approximation.
The FAM is nothing but a trick to perform this numerical
differentiation in the single-particle (Kohn-Sham) Hamiltonian $h[\rho]$.

The residual field
$\delta h(\omega)$ depends only on
the forward "ket" amplitudes $\ket{X_i(\omega)}$ and
backward "bra" ones $\bra{Y_i(\omega)}$.
In other words, it is independent of
bras $\bra{X_i(\omega)}$ and kets $\ket{Y_i(\omega)}$.
This is related to the fact that the transition density $\delta\rho(\omega)$
depends only on $\ket{X_i(\omega)}$ and $\bra{Y_i(\omega)}$.
\begin{equation}
\delta\rho(\omega) = \sum_i \left\{ \ket{X_i(\omega)}\bra{\phi_i}
                                  + \ket{\phi_i}\bra{Y_i(\omega)}
                            \right\} .
\end{equation}
We calculate the residual field by introducing a small real parameter
$\eta$ to realize the linear approximation \cite{NIY07}.
\begin{equation}
\delta h(\omega) = \frac{1}{\eta} 
\left( h\left[ \rho_\eta \right] -
 h_0 \right) ,
\label{FAM}
\end{equation}
where $h_0$ is the Hamiltonian for the ground state
and $\rho_\eta$ are defined by
\begin{equation}
\rho_\eta \equiv \sum_i \left\{
(\ket{\phi_i}+\eta\ket{X_i(\omega)})(\bra{\phi_i}+\eta\bra{Y_i(\omega)})
\right\} .
\end{equation}
Once $\ket{X_i(\omega)}$ and $\bra{Y_i(\omega)}$ are given,
the calculation of $h[\rho_\eta]$ is an easy task.
This does not require complicated programming, but only needs a small
modification in the calculation of $h[\rho]$.
Of course, eventually, we need to solve Eqs.~(\ref{RPAeqX}) and (\ref{RPAeqY})
to determine the forward and backward amplitudes.
We use an iterative algorithm to solve this problem.
Namely, we start from initial amplitudes $\ket{X_i^{(0)}}$ and
$\bra{Y_i^{(0)}}$, then update them in every iteration,
($\ket{X_i^{(n)}},\bra{Y_i^{(n)}}) \rightarrow
(\ket{X_i^{(n+1)}},\bra{Y_i^{(n+1)}})$,
until the convergence.
In each step, we calculate $\delta h(\omega)$ using the FAM
as Eq.~(\ref{FAM}).
For more details, readers are referred to the reference \cite{NIY07}.
\begin{figure}[t]
\begin{center}
\includegraphics[width=\textwidth]{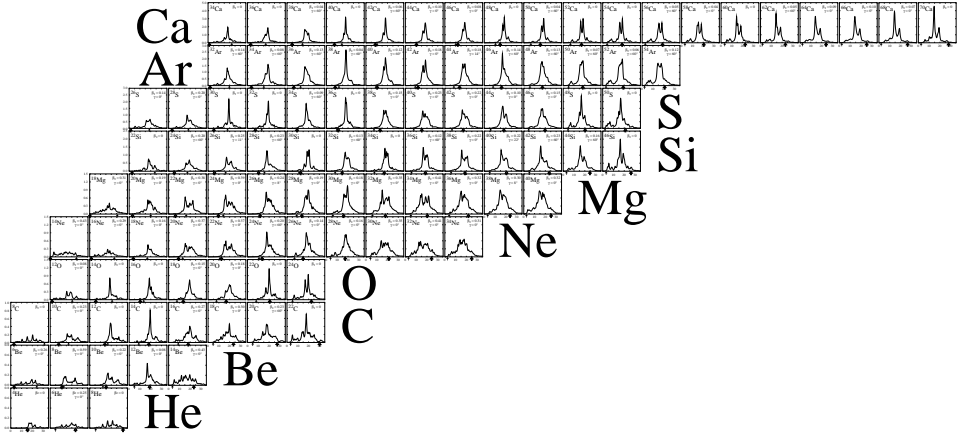}
\caption{Electric dipole strength distribution in light nuclei.
The horizontal axis corresponds to excitation energy of $0\sim 35$ MeV.
The SkM* parameter set and
the smoothing parameter of $\Gamma=1$ MeV is used.
}
\label{fig:Dipole_FAM}
\end{center}
\end{figure}

We have developed a parallelized computer program of the FAM
for a Skyrme functional
in the three-dimensional (3D) coordinate-space representation \cite{INY09}.
Currently, we are performing a systematic calculation of the electric
dipole response in even-even nuclei.
So far, we have calculated the photoabsorption cross section in nuclei
with $A\lesssim 100$.
In Fig.~\ref{fig:Dipole_FAM}, we demonstrate a part of our achievement
for nuclei up to Ca isotopes.

For nuclei with $A\leq 40$,
the observed strength up to 30 MeV exhausts only $60\sim 100$ \% of
the TRK sum-rule value \cite{HW01}.
This indicates that the considerable amount of the GDR strength is
located above 30 MeV in light nuclei.
We also observe that, although the RPA (FAM) calculation reproduces
a gross feature of the dipole strength distribution,
it systematically underestimates the GDR peak energy by a few MeV
for light nuclei \cite{INY09}.

\subsection{FAM with pairing correlations}
\label{sec:FAM_HFB}
\begin{figure}[t]
\begin{center}
\includegraphics[scale=0.5,angle=-90]{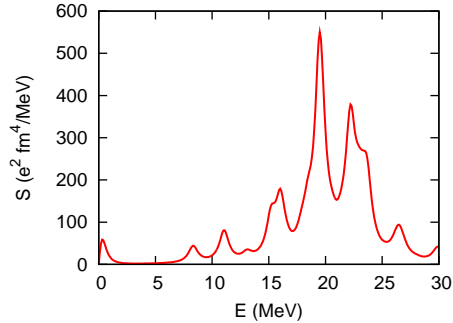}
\caption{Calculated isoscalar monopole strength distribution for $^{50}$Ca.
The SkM* parameter set and
the smoothing parameter of $\Gamma=1$ MeV is used.
}
\label{fig:50Ca}
\end{center}
\end{figure}
The FAM in the previous section can be extended to superfluid nuclei,
namely, to the QRPA with the HFB formalism.
A self-consistent solution of Eq.~(\ref{HFB}) 
determines the ground-state densities $(\rho_0,\kappa_0)$
and the ground-state Hamiltonians $(h_0,\Delta_0)$.
Then, following the same argument as that in Ref.~\cite{NIY07},
we can derive equations for the residual fields, $\delta h(\omega)$
and $\delta\Delta(\omega)$ as follows:
\begin{eqnarray}
\delta h(\omega) &=& \frac{1}{\eta} \left(
h[\rho_\eta,\kappa_\eta] - h_0 \right) , \\
\delta \Delta(\omega) &=& \frac{1}{\eta} \left(
\Delta[\rho_\eta,\kappa_\eta] - \Delta_0 \right) ,
\end{eqnarray}
where the density and pairing tensor $(\rho_\eta, \kappa_\eta)$ are defined by
\begin{eqnarray}
\rho_\eta &=& ( V^* + \eta U X ) (V + \eta U^* Y )^T ,\\
\kappa_\eta &=& ( V^* + \eta U X ) ( U + \eta V^* Y )^T .
\end{eqnarray}
Here, the forward and backward amplitudes $(X_{\mu\nu},Y_{\mu\nu})$
have subscripts $\mu\nu$ to specify two-quasiparticles.
On the other hand, the subscripts of $(U_{k\mu},V_{k\nu})$ indicate
a basis of the single-particle space ($k$) and the quasiparticle ($\mu$).
Again, utilizing an iterative algorithm for solution of the QRPA equation,
we can solve the QRPA linear-response equation
without explicitly calculating the residual interactions.

We show in Fig.~\ref{fig:50Ca} an example of our FAM calculation
for isoscalar monopole response in $^{50}$Ca.
We use the same parameter set and the same pairing energy functional
as those in Ref.~\cite{TE06}.
The quasiparticle states are truncated by the maximum quasiparticle energy
of $E_{\rm qp}=200$ MeV.
The result agrees with Fig.~1 in Ref.~\cite{TE06}.
The peak near zero energy should be associated with a small mixture of
the spurious mode (pairing rotation).

\section{Canonical-basis time-dependent HFB method}
\label{sec:Cb-TDHFB}
In Secs.~\ref{sec:QRPA} and \ref{sec:FAM},
we discuss methods to calculate linear response in nuclei,
based on the time-dependent density-functional theory.
In this section, we will show a feasible real-time method which
is, in principle, applicable to the non-linear regime as well.

The time-dependent Hartree-Fock (TDHF) method
in the 3D coordinate representation
is a well established method to study nuclear dynamics \cite{Neg82}.
However, it cannot describe particle-particle (hole-hole) pairing correlations.
The pairing correlations are supposed to be very important not only
for static properties but also for nuclear dynamics.
For instance, it is well known that the life time of spontaneous
fission is very different between even and odd nuclei, which is
supposed to be due to the pairing correlations.
A straightforward extension of the TDHF including the pairing correlations
is, of course,
the time-dependent Hartree-Fock-Bogoliubov (TDHFB) theory \cite{BR86}.
However it uses the quasi-particle orbitals instead of the occupied orbitals
whose number is, in principle, infinite.
Thus, the accurate calculation of TDHFB is presently impractical
and a new feasible approach is highly desirable.

In this section, we present the equations of motion
of ``Canonical-basis TDHFB'' (Cb-TDHFB) method
which we have developed recently \cite{Eba10}.
Then, we apply the method to the linear-response calculations using
the full Skyrme functional to show its reliability.
For more details, readers should be referred to the reference \cite{Eba10}.

\subsection{Basic equations}
\label{sec:Cb-TDHFB-eq}
Our starting point is that the TDHFB state can be written in the
canonical form as
\begin{equation}
\ket{\Psi(t)}=\prod_{k>0} \left\{
u_k(t) + v_k(t) c_k^\dagger(t) c_{\bar k}^\dagger(t) \right\} \ket{0} ,
\end{equation}
where the creation operator of particles
at the canonical state $\ket{\phi_k(t)}$ is expressed as
$\hat{c}_k^\dagger(t) = \sum_\sigma \int d\vec{r} \phi_k(\vec{r}\sigma; t)
 \hat{\psi}^\dagger(\vec{r}\sigma)$.
Here, the state $k$ and $\bar{k}$ are not necessarily related to each other
by the time reversal, and the time-dependent $(u,v)$ factors are complex
numbers.
Using the density matrix and pairing tensor appearing in the HFB equation
(\ref{HFB}), one can write $\rho_k(t)=|v_k(t)|^2$ and
$\kappa_k(t)=u_k^*(t) v_k(t)$ as
\begin{eqnarray}
\label{rho_k}
\rho_k(t)&=&
 \sum_{\mu\nu}
 \inproduct{\phi_k(t)}{\mu}\rho_{\mu\nu}(t)\inproduct{\nu}{\phi_k(t)}
 =\sum_{\mu\nu}
 \inproduct{\phi_{\bar k}(t)}{\mu}\rho_{\mu\nu}(t)
  \inproduct{\nu}{\phi_{\bar k}(t)} , \\
\label{kappa_k}
\kappa_k(t)&=&\sum_{\mu\nu} 
\inproduct{\phi_k(t)}{\mu}\inproduct{\phi_{\bar k}(t)}{\nu}
\kappa_{\mu\nu}(t) .
\end{eqnarray}
Then, utilizing the TDHFB equation, we obtain the following equations
for the time evolution of $\rho_k(t)$ and $\kappa_k(t)$.
\begin{eqnarray}
\label{drho_dt}
i\frac{d}{dt}\rho_k(t) &=&
\kappa_k(t) \Delta_k^*(t)
-\kappa_k^*(t) \Delta_k(t) , \\
\label{dkappa_dt}
i\frac{d}{dt}\kappa_k(t) &=&
\left(
\eta_k(t)+\eta_{\bar k}(t)
\right) \kappa_k(t) +
\Delta_k(t) \left( 2\rho_k(t) -1 \right) ,
\end{eqnarray}
where
\begin{eqnarray}
\label{Delta_k}
\Delta_k(t) &\equiv& -\sum_{\mu\nu}
\Delta_{\mu\nu}(t)
 \inproduct{\phi_k(t)}{\mu}\inproduct{\phi_{\bar k}(t)}{\nu} , \\
\eta_k(t) &\equiv& \bra{\phi_k(t)}h(t)\ket{\phi_k(t)}
+i\inproduct{\frac{\partial\phi_k}{\partial t}}{\phi_k(t)} .
\end{eqnarray}
So far, there is no approximation in addition to the TDHFB
is involved.
Now, we need to introduce an approximation for the pair potential.
Namely, the pair potential is assumed to be diagonal in the canonical basis.
\begin{equation}
\label{Delta_mn}
\Delta_{\mu\nu}(t) = -\sum_{k>0}
\Delta_k(t) \left\{
\inproduct{\mu}{\phi_k(t)}\inproduct{\nu}{\phi_{\bar k}(t)} -
\inproduct{\nu}{\phi_k(t)}\inproduct{\mu}{\phi_{\bar k}(t)} 
\right\} .
\end{equation}
In the static limit, this is identical to the BCS approximation.
With the approximation of Eq. (\ref{Delta_mn}),
one can derive the following simple equations for the time-dependent
canonical states.
\begin{equation}
\label{dphi_dt}
i\frac{\partial}{\partial t} \ket{\phi_k(t)} =
(h(t)-\eta_k(t))\ket{\phi_k(t)} , \quad\quad
i\frac{\partial}{\partial t} \ket{\phi_{\bar k}(t)} =
(h(t)-\eta_{\bar k}(t))\ket{\phi_{\bar k}(t)} .
\end{equation}

In summary,
the Cb-TDHFB equations consists of Eqs.
(\ref{dphi_dt}),
(\ref{drho_dt}), and (\ref{dkappa_dt}).
To derive these equations from the TDHFB equations,
we have assumed the diagonal property of the pair
potential, Eq. (\ref{Delta_mn}).

\begin{figure}[t]
\begin{center}
\includegraphics[scale=0.3,angle=-90]{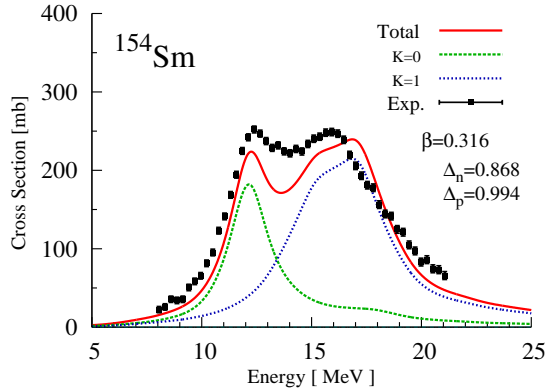}
\caption{Calculated (lines) and experimental (symbols) photoabsorption cross
section in $^{154}$Sm.
The SkM* parameter set and
the smoothing parameter of $\Gamma=2$ MeV is used.
See text for details.
}
\label{fig:154Sm_CbTDHFB}
\end{center}
\end{figure}
\subsection{Linear response calculation in real time}
\label{sec:Cb-TDHFB-cal}
We have applied the Cb-TDHFB method to study of electric dipole resonances
in Ne and Mg isotopes \cite{Eba10}.
Here, we apply the method to GDR in the deformed $^{154}$Sm nucleus.
We calculate the time evolution of the electric dipole moment, starting
from the HF+BCS ground state with a perturbative
instantaneous external dipole field.
Then, we perform the Fourier transform to obtain the response function.
The details of the calculation can be found in Ref.~\cite{Eba10}.

We show in Fig.~\ref{fig:154Sm_CbTDHFB} the calculated photoabsorption
cross section in $^{154}$Sm.
Although the pair potential is simplified in the Cb-TDHFB calculation,
the result is almost identical to the QRPA calculation shown in
Fig.~\ref{fig:154Sm_QRPA},
except for a small difference seen in the second peak.
We have examined the origin of this difference and found that
the neglect of the residual Coulomb in the QRPA calculation
is responsible for this small discrepancy.
Thus, we may conclude that the Cb-TDHFB can reproduce the
QRPA result at its small amplitude limit.

It should be emphasized that the computational cost of the Cb-TDHFB
is significantly smaller than the QRPA.
The present calculation in the full 3D space
can be achieved in roughly 50 CPU hours,
while the QRPA calculation in Fig.~\ref{fig:154Sm_QRPA}, that is
restricted to the axially symmetric nuclei,
requires roughly 1,000 CPU hours.
This is because the Cb-TDHFB treats only the canonical states whose number
is the same order as the particle number.
In contrast, in the QRPA (or in the TDHFB), we need to treat the
quasiparticle states whose number is the same as the dimension of
the model space.

\section{Conclusion}
We have presented our recent developments for studies of nuclear
response functions.
The parallelized quasiparticle random-phase-approximation (QRPA) code is
now ready for investigation for heavy axially deformed nuclei.
The finite-amplitude method (FAM) was applied to systematic
investigation of the photoabsorption cross section in light nuclei.
Recently, the QRPA version of the FAM has been developed
for superfluid nuclei, including the pairing correlations.
We also presented the canonical-basis formulation of the TDHFB.
This is applicable to large-amplitude nuclear dynamics beyond
the linear approximation.

The work is supported by Grant-in-Aid for Scientific Research
(Nos. 21340073 and 20105003).
Computational resources were provided by the Joint Research Program at
Center for Computational Sciences, University of Tsukuba,
and by the Large Scale Simulation Program
of High Energy Accelerator Research Organization (KEK).
The numerical calculations were also performed
on RIKEN Integrated Cluster of Clusters (RICC).
We would like to thank the JSPS Core-to-Core Program ``International
Research Network for Exotic Femto Systems'' and the UNEDF SciDAC
collaboration under DOE grant DE-FC02-07ER41457.

\bibliographystyle{unsrt}
\bibliography{nuclear_physics,myself}

\end{document}